\documentclass[fleqn,11pt]{wlscirep}
\usepackage[utf8]{inputenc}
\usepackage[T1]{fontenc}
\usepackage{bm}
\usepackage{mathrsfs}
\usepackage{enumitem}
\usepackage{bbold}
\usepackage{mathtools}

\usepackage{braket}

\title{Degenerate topological line surface phonons in quasi-1D double helix crystal SnIP}

\author[1]{Bo Peng}
\author[2,3]{Shuichi Murakami}
\author[1,4,*]{Bartomeu Monserrat}
\author[2,3,${\dagger}$]{Tiantian Zhang}

\affil[1]{TCM Group, Cavendish Laboratory, University of Cambridge,
J. J. Thomson Avenue, Cambridge CB3 0HE, United Kingdom}

\affil[2]{Department of Physics, Tokyo Institute of Technology, Ookayama, Meguro-ku, Tokyo 152-8551, Japan}

\affil[3]{Tokodai Institute for Element Strategy, Tokyo Institute of Technology, Nagatsuta, Midori-ku, Yokohama, Kanagawa 226-8503, Japan}

\affil[4]{Department of Materials Science and Metallurgy, University of Cambridge, 27 Charles Babbage Road, Cambridge CB3 0FS, United Kingdom}

\affil[*]{bm418@cam.ac.uk}

\affil[${\dagger}$]{ttzhang@stat.phys.titech.ac.jp}

\begin{abstract}
Degenerate points/lines in the band structures of crystals have become a staple of the growing number of topological materials. The bulk-boundary correspondence provides a relation between bulk topology and surface states. While line degeneracies of bulk excitations have been extensively characterized, line degeneracies of surface states are not well understood. We show that SnIP, a quasi-one-dimensional van der Waals material with a double helix crystal structure, exhibits topological nodal rings/lines in both the bulk phonon modes and their corresponding surface states. Using a combination of first-principles calculations, symmetry-based indicator theories and Zak phase analysis, we find that two neighbouring bulk nodal rings form doubly degenerate lines in their drumhead-like surface states, which are protected by the combination of time-reversal symmetry $\mathcal{T}$ and glide mirror symmetry $\bar{M}_y$. Our results indicate that surface degeneracies can be generically protected by symmetries such as $\mathcal{T}\bar{M}_y$, and phonons provide an ideal platform to explore such degeneracies.
\end{abstract}
\begin{document}
\flushbottom
\maketitle

\thispagestyle{empty}

\section*{Introduction}

% Bulk band degeneracy

Degeneracies in the bulk energy bands of crystals were intensively studied in the early days of band theory \cite{Herring1937}. Recent breakthroughs have shed new light into this old topic by associating topological invariants with these degenerate points or lines \cite{Heikkila2011,Weng2016,Bernevig2018}, and by extending the discussion from electrons to phonons \cite{Stenull2016,Liu2017a,He2018}, photons \cite{Jia2019}, magnons \cite{Li2016d}, and excitons \cite{Wu2017a}. Among the degenerate points, twofold degenerate Weyl points always appear in pairs that can be topologically characterized with opposite Chern numbers \cite{Xu2015,Lv2015,Weng2015,Yang2015}, and fourfold degenerate Dirac points have a linear dispersion in both electronic \cite{Novoselov2004,Zhang2005,Wang2012a,Liu2014b,Armitage2018,Peng2018d} and phononic systems \cite{Zhang2018a,Li2021}. When the band crossings are one-dimensional in momentum space, they can form nodal lines, nodal rings or nodal chains, depending on their shape \cite{Weng2015b,Fang2015,Yu2015b,Bzdusek2016,Huang2016b,Hu2016a,Bian2016,Yang2018,Lian2019}, and these line crossings are protected by symmetries such as mirror or $\mathcal{PT}$ (where $\mathcal{P}$ is the spatial inversion symmetry and $\mathcal{T}$ is the time reversal symmetry) \cite{Fang2016,Peng2019}. Recent advances, using both group theoretical analysis and high-throughput calculations, have enabled a comprehensive understanding of bulk band degeneracies in both electrons \cite{Zhang2019a,Tang2019,Vergniory2019} and phonons \cite{Li2021}. These topological degeneracies exhibit unique physical properties such as a `quantum anomaly' \cite{Son2013,Xiong2015}, enabling various applications from spintronics \cite{Belopolski2019} to topological quantum computation \cite{Bouhon2020}.

%However, although degeneracies in the bulk energy bands are now well understood, using both group theoretical analysis and high-throughput calculations \cite{Zhang2019a,Tang2019,Vergniory2019}, degeneracies in the surface states remain largely unexplored. 

% Topological semimetals with point band crossings include Dirac and Weyl semimetals , while those with line band crossings are called nodal-line, nodal-ring, or nodal-chain semimetals \cite{Weng2015b}. As a precursor state for other topological phases, nodal-line semimetal, for instance, can be driven into Dirac semimetal in Cu$_3$PdN \cite{Yu2015b}, Weyl semimetal in TaAs \cite{Weng2015}, and topological insulator in three-dimensional graphene networks \cite{Weng2015c} by spin-orbit coupling. On the other hand, nodal-line semimetal itself can exist in the presence of strong spin-orbit coupling \cite{Bian2016}, which exhibits exotic properties such as topological surface flat bands for high-temperature superconducting applications \cite{Burkov2011,Fang2015}. 

\begin{figure}[h]
\centering
\includegraphics[width=0.5\linewidth]{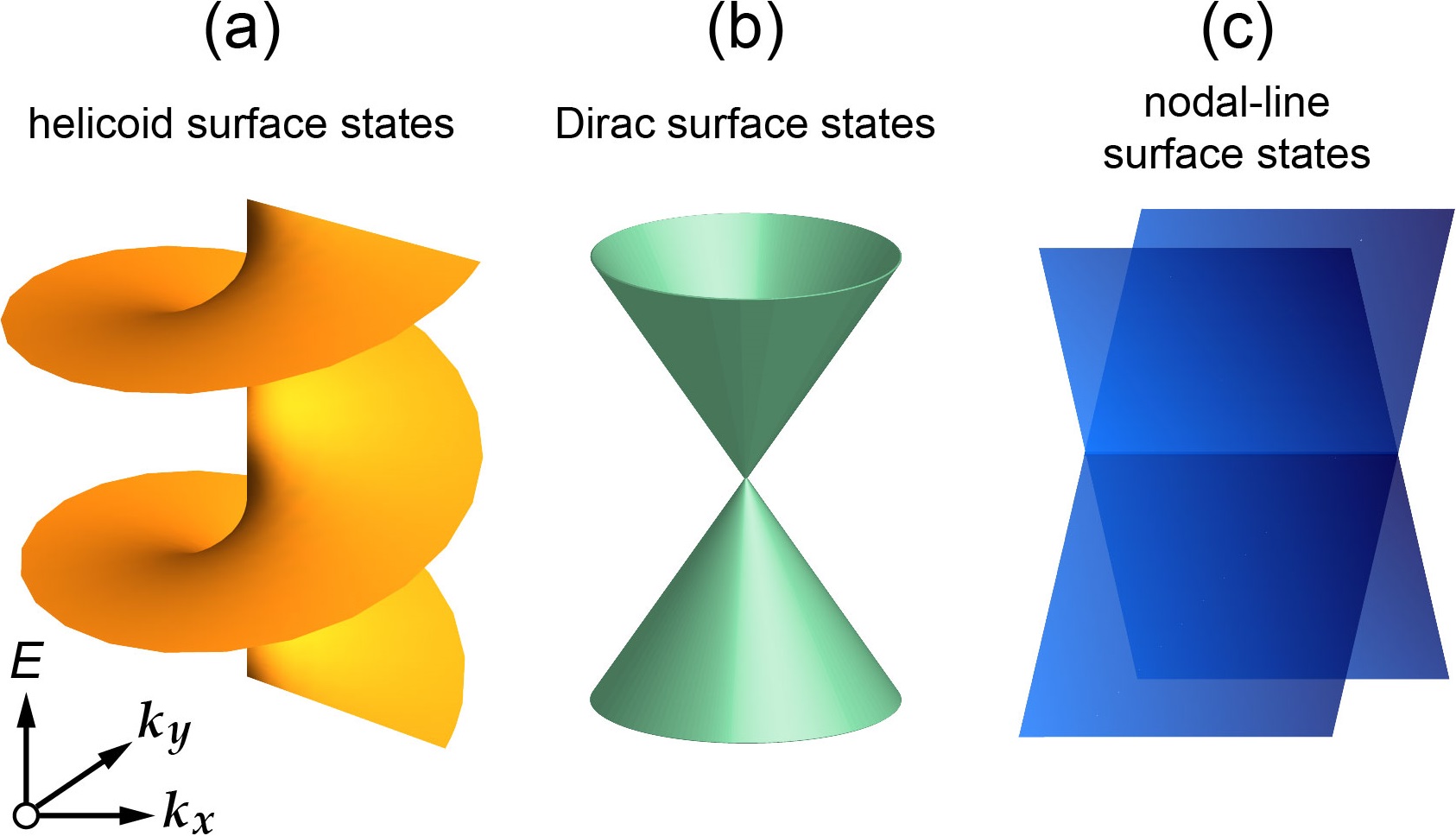}
\caption{Different types of the surface states in topological materials. (a) Non-degenerate helicoid surface states, (b) zero-dimensional degenerate Dirac surface states, and (c) one-dimensional degenerate nodal-line surface states.}
\label{schematic} 
\end{figure}

% Surface states degeneracy
As a result of the bulk-boundary correspondence, the degenerate bulk points and lines are associated with topologically protected surface states, including surface arcs associated with Weyl and Dirac points \cite{Sun2015,Huang2015b,Deng2016,Tamai2016} and drumhead-like flat bands associated with nodal lines \cite{Burkov2011,Fang2015}. Similar to bulk band degeneracies, surface states can also exhibit band crossings in both semi-infinite slabs and finite slabs (with top and bottom surfaces) \cite{Okugawa2014}. Here we focus on semi-infinite systems with only one surface. For Weyl points with Chern numbers of $\pm 1$, the associated helicoid surface states, whose isoenergy contours are Fermi arcs, form a continuous 2D surface in the energy-momentum space $E$-\textbf{k} and have no degeneracy \cite{Fang2016a}, as schematically shown in Figure~\ref{schematic}(a). In topological insulators and topological crystalline insulators, the surface states become degenerate at a single point \cite{Fu2007,Zhang2009,Chen2009,Dziawa2012,Wojek2015}, forming a surface Dirac cone, as shown schematically in Figure~\ref{schematic}(b). In principle, two surface states can also cross along a line in the surface Brillouin zone, as shown schematically in Figure~\ref{schematic}(c). However, like the bulk case, these surface line crossings need to be protected by additional symmetries \cite{Fang2016a,Wang2016k,Alexandradinata2016,Ezawa2016,Kuo2019}. While nodal line degeneracies in the bulk states have been thoroughly explored in a variety of quasiparticle band structures, the degenerate lines in the surface states have received less attention. The few examples of surface nodal lines all correspond to electronic band structures \cite{Fang2016a,Wang2016k,Alexandradinata2016,Ezawa2016,Hosen2018,Kuo2019} or artificial systems \cite{Cheng2020,Cai2020}. It is therefore desirable to identify more candidates in other quasiparticle spectra to fully understand the role of surface degeneracies.
% \sout{When two surface states form a degenerate line [Fig.~\ref{schematic}(c)], extra crystal symmetry such as glide mirror symmetry is required \cite{Fang2016a,Kuo2019}, because with only time-reversal symmetry and non-symmorphic symmetry, the surface states can only be degenerate at isolated points. This is similar to the case of degenerate nodal lines/rings in the bulk bands, where extra mirror symmetry or inversion symmetry should be preserved \cite{Fang2016,Peng2019}.} 

Among all the quasiparticles, phonons are of particular interest as they describe intrinsic ionic vibrations. Moreover, the bosonic nature of phonons facilitates the exploration of topological phenomena in their entire spectrum. This is in contrast to fermonic systems (\textit{e.g.} electronic systems) in which there is an additional constraint that degeneracies need to occur near the Fermi level to have an impact on the behaviour of the material. Furthermore, time-reversal symmetry is hard to break in phononic systems, and consequently this extra symmetry requirement is automatically satisfied in phonons. These advantages have led to a recent surge of interest in characterising topological properties of phonons \cite{Stenull2016,Liu2017a,He2018,Zhang2018a,Miao2018,Li2018a,Xia2019,Zhang2019c,Liu2020,Li2021,Peng2020a,Liu2020b,Wang2021,Tang2021,Liu2021,Liu2021a,Liu2021b,You2021,Xie2021,Zheng2021,Wang2021a,Peng2021}, but degenerate lines in surface phonon states have not yet been reported. 

In this work, we propose that SnIP, an inorganic semiconductor with nodal-ring phonons in the bulk states, exhibits nodal-line surface states in the phonon spectrum using a combination of first-principles calculations, group theory analysis and Zak phase analysis. These surface states are doubly degenerate in a nodal line on the (100) surface. The surface nodal line is protected by the anti-unitary symmetry $\vartheta=\mathcal{T}\bar{M}_y$, which is a combination of time-reversal symmetry $\mathcal{T}$ and glide mirror symmetry $\bar{M}_y$. With different surface terminations, we can alternatively get two surface nodal lines on different parts of the (100) surface Brillouin zone, which can be understood by different Zak phases in the presence of $\mathcal{T}\bar{M}_y$ symmetry. Our results suggest that similar surface degenerate lines/points will be found in other materials when there are extra symmetries to protect these surface degeneracies, and we believe that phonons can be a primary platform to study these degeneracies because of the presence of time-reversal symmetry and the flexibility to study the degeneracies in the entire phonon spectrum.

% for a system with the mirror symmetry $k_x = k_y$, when there is a band crossing on the mirror plane, the crossing bands satisfy $E_{i}(k)=E_{i+1}(k)$ ($i$ is the band index). Because $k$ has three variables ($k_x$, $k_y$, $k_z$) to satisfy two equations, \textit{i.e.}, $k_x = k_y$ and $E_{i}(k)=E_{i+1}(k)$, the solution space must be one-dimensional, which restricts the band crossing to a 1D continuous ring/line on the mirror-invariant plane \cite{Fang2016}. 

% Russian journal: It has been recognized that collective vibrational modes in diverse biomacromolecules on subpicosecond time scales can play a critical role in controlling conformational changes and triggering biochemical reactions and energy transport

% SSH model? Electronic system, hard! Phonons!

\section*{Results}

\subsection*{Crystal structure and phonon dispersion}

SnIP is an inorganic semiconductor with a double helix crystal structure, first synthesized in 2016, and can be formed without any templates \cite{Pfister2016,Saleh2016,Utrap2017}. The strong covalent bonds along the helices and the weak van der Waals interactions across the helices combine the electronic properties of inorganic semiconductors, such as high electron mobility \cite{Li2017b}, and the mechanical properties of polymers, such as high flexibility \cite{Ott2019}. In addition, by atomic substitution, a large material family with double helix structures has been predicted using first-principles calculations, providing a wide range of band gaps between 1 and 2.5 eV \cite{Baumgartner2017}. With a tunable band gap and high mechanical flexibility, these quasi-one-dimensional semiconductors are extremely promising for next generation devices ranging from mechanical sensors to optoelectronics \cite{Ott2019,Baumgartner2017}. Moreover, they can serve as a material platform to realize one-dimensional physical models such as the Su-Schrieffer-Heeger model \cite{Su1980} and Majorana quantum wires \cite{Kitaev2001,Franz2013}.

\begin{figure*}[h]
\centering
\includegraphics[width=\linewidth]{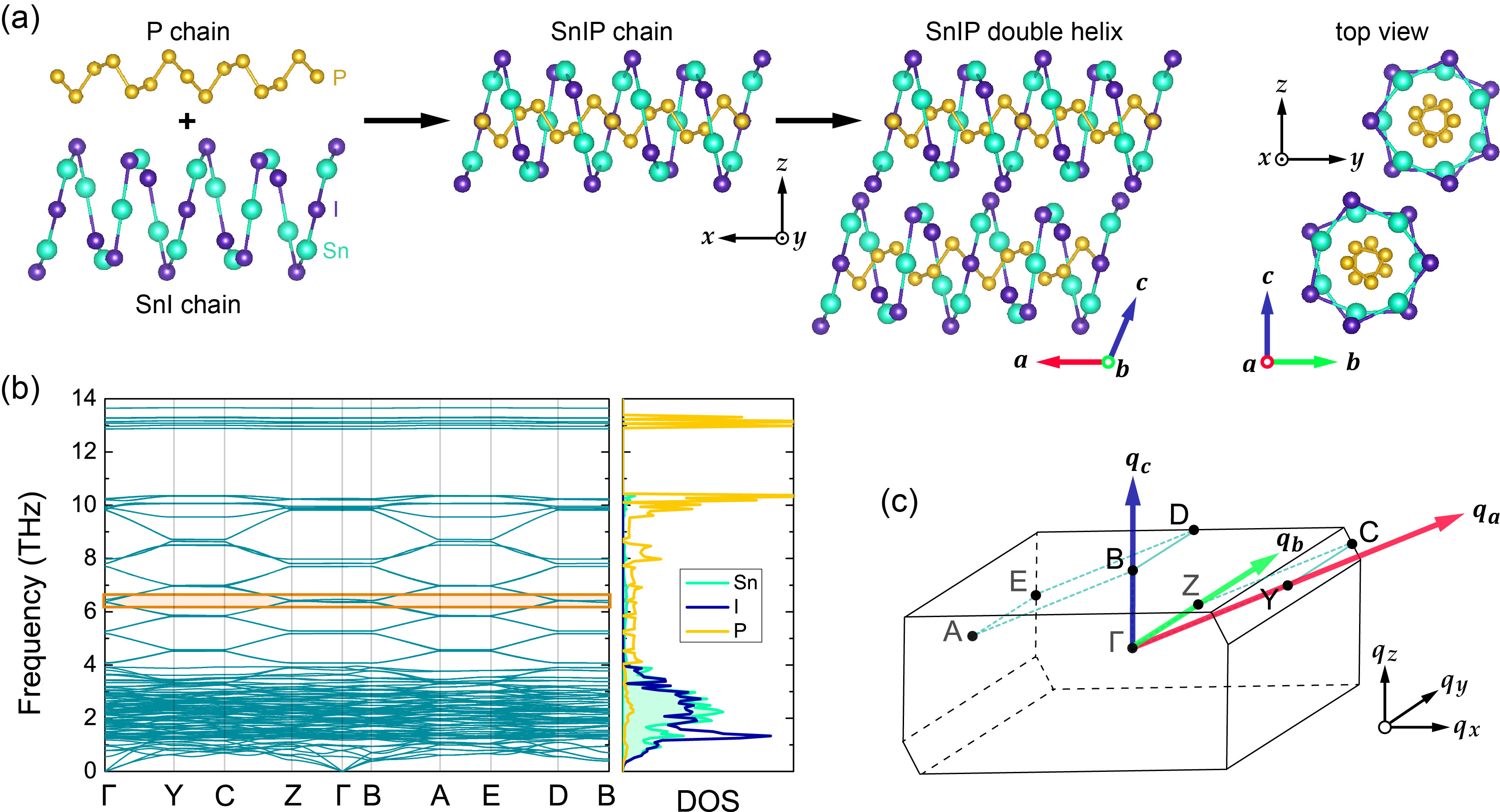}
\caption{Structural and vibrational properties of SnIP. (a) Crystal structure, (b) phonon dispersion and (c) Brillouin zone of SnIP. The orange region in the phonon dispersion highlighted in (b) corresponds to the topological features that we focus on. The phonon DOS is also shown in (b).}
\label{crystal} 
\end{figure*}

Bulk SnIP crystallizes in the monoclinic space group $P2/c$ (No.\,13) with two double helices (SnIP)$_7$ in the unit cell with stoichiometry (SnIP)$_{14}$. The two double helices, alternatively left- and right-hand twisted, are stacked along the $a$ direction. Each double helix is formed by an inner helical P chain and an outer helical SnI chain, as shown in Figure~\ref{crystal}(a). The (SnIP)$_7$ unit in each SnIP chain winding results in a 7/2 helix. From one phosphorus (yellow) atom to the next, the turning angle is about $\pm 360^{\circ}$/7 followed by a translation of $c$/7, and the same holds true for the tin (cyan) and iodine (navy) atoms in the SnI helix \cite{Muller2017}. The calculated lattice parameters of $a$ = 7.889 \AA, $b$ = 9.768 \AA, $c$ = 18.422 \AA, and monoclinic distortion $\beta$ = 110.21$^{\circ}$ match well with the measured ones \cite{Pfister2016,Utrap2017}.

The calculated phonon dispersion of SnIP is presented in Figure~\ref{crystal}(b). It exhibits wide bandwiths along the helical direction, corresonding to the $\Gamma$-Y, C-Z, B-A and E-D high-symmetry lines in Figure~\ref{crystal}(c), but, by contrast, phonon bands along other high-symmetry lines, especially those perpendicular to the helices, have a relatively flat dispersion, a result of the quasi-one-dimensional nature of this material. This is consistent with the fact that SnIP has strong covalent bonds along the one-dimensional helices but only weak van der Waals-like interactions between the double helices. As a result, SnIP fibers can be macroscopically bent by up to 90$^{\circ}$ without any significant Raman mode shifts, indicating that SnIP is an ultrasoft inorganic material and bending the helices has negligible influence on the phononic properties \cite{Ott2019}. 

There are 42 atoms in the unit cell of SnIP, corresponding to 126 phonon branches. The calculated atom-projected phonon density of states (DOS) in Figure~\ref{crystal}(b) indicates that the low frequency modes below 4 THz are mainly vibrations involving the heavier Sn and I atoms, while those between 4-14 THz are dominated by the motion of the lighter P atoms. There are 14 phosphorus atoms in the unit cell, leading to 42 P-dominated bands in the phonon spectra, spanning branches 85 to 126. The phonon modes from 4 to 10 THz (branches 85-108) are more dispersive compared to the almost flat bands between 10 and 14 THz. %The vibrations in the dispersive region mostly correspond to intrachain motion involving strong covalent bonds, whereas those in the flatter region mostly correspond to interchain vibrations across the helices with much weaker interactions. 
The two double helices are related to each other by inversion/glide mirror symmetry, so there always exist two bands with similar eigenvalues and similar eigenvectors, corresponding to the P atoms from the two separate P chains, respectively (see the phonon dispersion between 4-8.5 THz in Supplementary Note 1 of the Supplementary Information).

Hereafter we focus on the frequency range 4-10 THz because the corresponding phonon bands are relatively `clean' due to the large mass difference between phosphorus and the other two elements that separates them from the lower energy bands, and due to the strong intrachain vibrations that make them highly dispersive. This makes them an ideal playground for exploring topological features. In addition, these modes in the range 4-10 THz contribute to nearly half the thermal conductivity along the helix direction (see the phonon transport properties in Supplementary Note 2 of the Supplementary Information).

% The highly dispersive modes along the helices have much higher group velocities and can propagate heat more effectively. It is therefore worth investigating the thermal transport properties of SnIP.

%These 42 bands are classified into three groups according to the three degrees of freedom in the real space. For the first group from 4 to 8 THz, the phonon modes are more dispersive compared to the almost flat bands between 8 and 14 THz. The vibrations in the first region are along the helices with strong covalent bonds, whereas the other two groups are vibrations across the helices. Because the two double helices are related to each other by symmetries such as glide mirror or inversion symmetry, there are always two bands with similar eigenvalues and similar eigenvectors (see the phonon dispersion between 4-8 THz in the Supplementary Material).

\subsection*{Nodal ring phonons in SnIP}

\begin{figure*}[h]
\centering
\includegraphics[width=\linewidth]{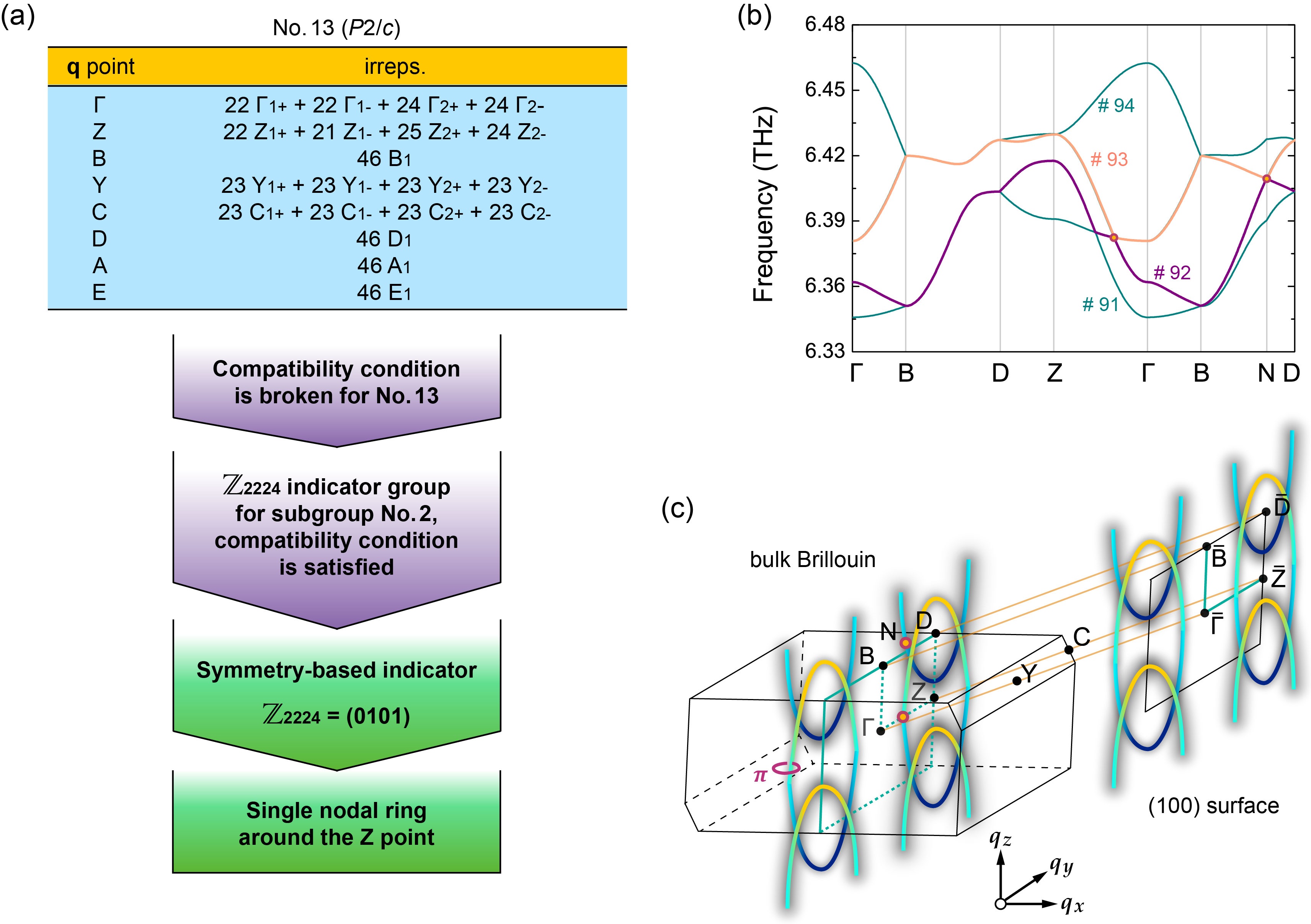}
\caption{Topological properties of SnIP. (a) Diagnosis process for the single nodal ring around the Z point using symmetry-based indicator theory for the lowest 92 bands. (b) Phonon band crossings between branches 92 and 93, and (c) the nodal rings in the bulk Brillouin zone, as well as their projections on the (100) surface.
}
\label{ring} 
\end{figure*}

For the phonon modes between 4 and 10 THz, dominated by the vibrations of the relatively light P atoms, the phonon bands have a large dispersion along the helical direction but are almost flat along the perpendicular directions, as shown in Figure~\ref{crystal}(b). The band folding of the seven P atoms in a single helical P chain \cite{Muller2017} leads to several band crossings near the Brillouin zone boundary, especially in the flat band regions. After a comprehensive search for band crossing points between bands 85-108 (\textit{i.e.}, 4-10 THz), we find that the most promising band crossings are formed by phonon branches 92 and 93 with clean topological features in both bulk and surface states. Marked with \#~92 and \#~93 in Figure~\ref{ring}(b), these two bands cross along the $\Gamma$-Z high-symmetry line. 

To understand the topological information of the band crossings between branches~92 and~93, we use symmetry-based indicator theories to analyze the degeneracies \cite{Po2017,Song2018b,Kruthoff2017,Zhang2019c,Zhang2020,Zhang2021}. Because of the existence of a band crossing along the $\Gamma$-Z high-symmetry line, the compatibility condition is broken for the lowest 92 bands, as the compatibility relations can only be satisfied when the band structures have no (non-accidental) band crossings along any high-symmetry lines or at any high-symmetry points \cite{Song2018b}. So we cannot use the symmetry-based indicator for space group No.\,13. Instead, we need to find a maximal subgroup of SnIP that simultaneously satisfies the compatibility condition and has a nontrivial symmetry-based indicator. Among all the subgroups of space group No.\,13 (space groups No.\,2, No.\,3 and No.\,7), No.\,2 is the only subgroup which both satisfies the compatibility condition and has a symmetry-based indicator group: (i) the compatibility condition is satisfied because the twofold rotation symmetry in No.\,13 that protects the band crossing along $\Gamma$-$Z$ is absent in No.\,2; (ii) space group No.\,2 has a non-trivial symmetry-based indicator of $\mathbb{Z}_{2224} \equiv \mathbb{Z}_{2} \times \mathbb{Z}_{2} \times \mathbb{Z}_{2} \times \mathbb{Z}_{4}$, as discussed in detail in Ref.\,[\citenum{Song2018b}]. Therefore we can use $\mathbb{Z}_{2224}$ as indicators to diagnose the topological degeneracy. For space group No.\,2, there are eight high-symmetry points, \textit{i.e.} high-symmetry points ($q_a,q_b,q_c$) with $q_{a,b,c} = 0,0.5$. We obtain the irreducible representations (irreps.) at the eight corresponding high-symmetry points $\Gamma$ (0,0,0), Z (0,0.5,0), B (0,0,0.5), Y (0.5,0,0), C (0.5,0.5,0), D (0,0.5,0.5), A (0.5,0,0.5), and E (0.5,0.5,0.5), as shown in Figure~\ref{ring}(a). We then compute the indicator accordingly. The resulting indicator (0101) implies that there is a single nodal ring around the Z point in SnIP formed by branches~92 and~93, which is verified by the numerical calculations in Figure~\ref{ring}(c).

% We obtain a symmetry-based indicator $\mathbb{Z}_{2224}$ of (0101), indicating a single nodal ring around the Z point.

We next examine in detail the nodal rings formed by bands 92 and 93. We zoom in the phonon dispersion on the $\Gamma$-B-D-Z plane in Figure~\ref{ring}(b). The band crossing formed by bands 92 and 93 along the $\Gamma$-Z high-symmetry line belongs to the single nodal ring around the Z point, % \sout{The degenerate rings are near the $\Gamma$-B-D-Z plane, crossing the $\Gamma$-Z high-symmetry line.}
carrying a quantized $\pi$ Berry phase. The nodal ring crosses through the Brillouin zone boundary around the B-D high-symmetry line, and consequently there are two nodal rings from two neighbouring Brillouin zones formed around the D point. We also plot the phonon dispersion around N $(-0.01,0.35,0.51)$, another band crossing point on the neighbouring nodal ring, as shown in Figure~\ref{ring}(b). Most interestingly, we find that the two neighbouring nodal rings projected on the (100) surface have a large overlap, and the two drumhead-like surface states of the two bulk nodal rings form a doubly degenerate line along $\bar{\textrm{B}}$-$\bar{\textrm{D}}$ (as discussed below). %\sout{ This doubly degenerate projection of the bulk Dirac cones is necessary for the formation of nodal-line surface states.}

\subsection*{Topological surface states of SnIP}

\begin{figure*}[h]
\centering
\includegraphics[width=\linewidth]{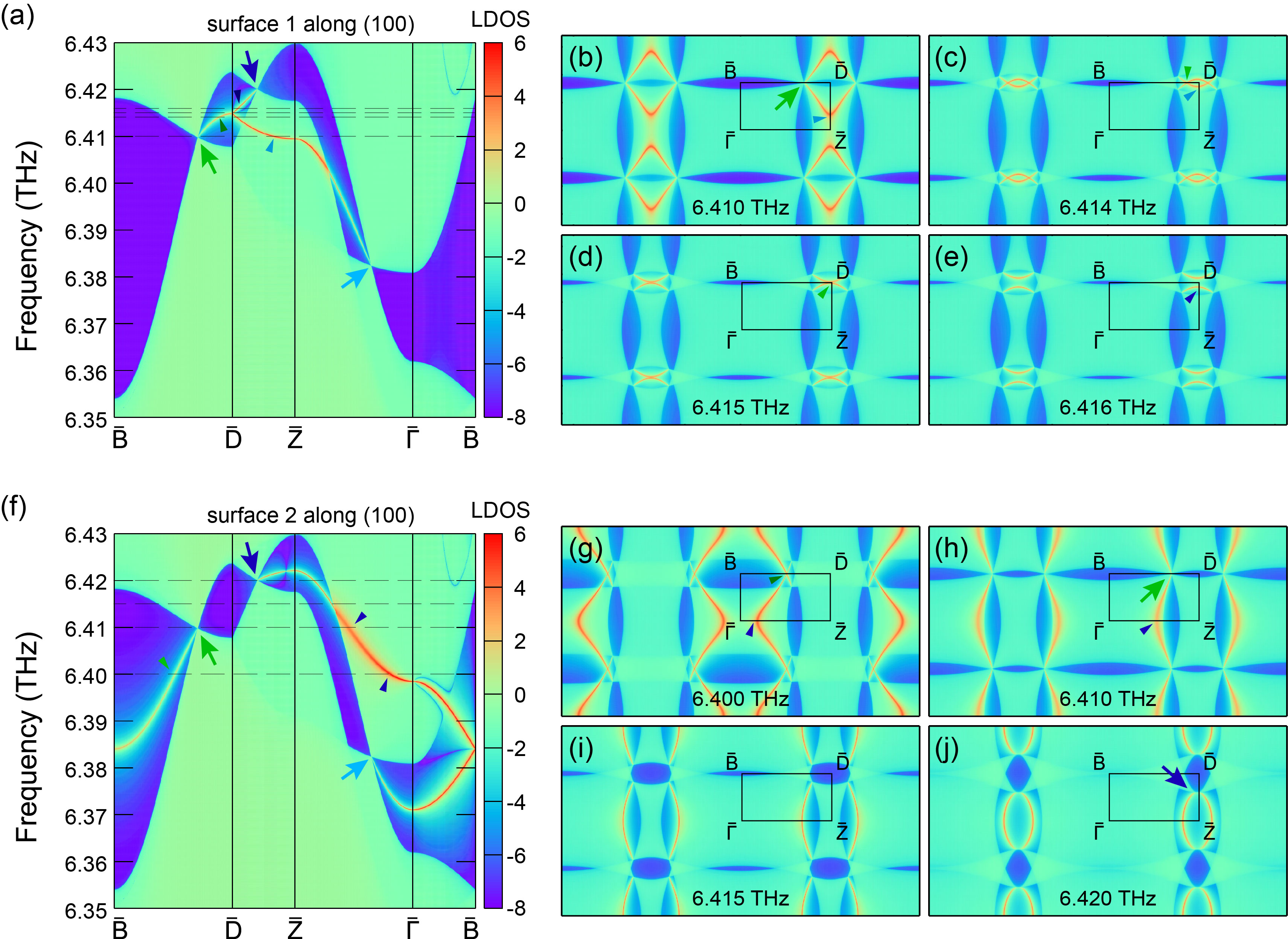}
\caption{Topological surface states of SnIP. (a) Topological surface states of surface 1 along the (100) direction, as well as the surface arcs for phonon frequencies of (b) 6.410 THz, (c) 6.414 THz, (d) 6.415 THz, and (e) 6.416 THz. (f) Topological surface states of surface 2 along the (100) direction, as well as the surface arcs for phonon frequencies of (g) 6.400 THz, (h) 6.410 THz, (i) 6.415 THz, and (j) 6.420 THz.}
\label{surface} 
\end{figure*}

We demonstrate the nodal-line surface states by calculating the surface local density of states (LDOS) from the imaginary part of the surface Green's function \cite{Wu2018}. We first study the surface along the (100) direction with the termination denoted as surface 1 in Figure~\ref{surface}(a)-(e). As shown in Figure~\ref{surface}(a), along the $\bar{\textrm{D}}$-$\bar{\textrm{Z}}$ and $\bar{\textrm{Z}}$-$\bar{\Gamma}$ high-symmetry lines, two separate projections of the bulk band crossings (marked by navy and cyan arrows) belong to two single nodal rings from two neighbouring Brillouin zones respectively. The surface states are within the projections of the two bulk nodal rings, forming two non-degenerate drumhead-like surface states marked by navy and cyan triangles in Figure~\ref{surface}(a) respectively. The surface states from the two crossing points along $\bar{\textrm{D}}$-$\bar{\textrm{Z}}$ and $\bar{\textrm{Z}}$-$\bar{\Gamma}$ merge with each other at $\bar{\textrm{D}}$, and result in a doubly degenerate surface nodal line along $\bar{\textrm{B}}$-$\bar{\textrm{D}}$ (green triangle). The surface nodal line ends at the crossing point of the two projected nodal rings marked by green arrow in Figure~\ref{surface}(a).

To have a better view of the doubly degenerate surface states along the $\bar{\textrm{B}}$-$\bar{\textrm{D}}$ high-symmetry line, we plot the isofrequency surface arcs for different phonon frequencies in Figure~\ref{surface}(b)-(e). As shown in Figure~\ref{surface}(b), at the exact frequency of the doubly degenerate projected bulk band crossings (green arrow) along $\bar{\textrm{B}}$-$\bar{\textrm{D}}$ at 6.410 THz, the two surface arcs (cyan triangle) merge at the projected bulk band crossings. For a phonon frequency of 6.414 THz, the two surface arcs (cyan triangle) move towards each other, forming two arc crossing points (green triangle) along $\bar{\textrm{B}}$-$\bar{\textrm{D}}$ in Figure~\ref{surface}(c). At 6.415 THz, the two arc crossing points move to the $\bar{\textrm{D}}$ point until they merge into a single arc crossing point [green triangle in Figure~\ref{surface}(d)]. Further increasing the phonon frequency eventually leads to two separate surface arcs as the two arcs no longer touch [navy triangle in Figure~\ref{surface}(f)]. The arc crossing points on the isofrequency plane belong to the doubly degenerate surface nodal line in the frequency-momentum space [green triangle in Figure~\ref{surface}(a)]. 

The surface nodal line in SnIP is protected by the combination of time-reversal symmetry and glide mirror symmetry $\vartheta=\mathcal{T}\bar{M}_y$. For systems with time-reversal symmetry $\mathcal{T}$ and any nonsymmorphic spatial symmetry $G$, the `Kramers-like' degenerate nodal line appears when $(\mathcal{T}G)^2 = -1$, either along the $\mathcal{T}G$ invariant lines or on the $\mathcal{T}G$ invariant planes, and a degenerate line can thus be protected by the $\mathcal{T}G$ symmetry. This applies to line degeneracies in both the bulk states \cite{Takahashi2017} and the surface states \cite{Wang2016k,Alexandradinata2016,Ezawa2016}. SnIP has glide mirror symmetry $\bar{M}_y= \{ M_y \ | \ 00\frac{1}{2} \}$, and $(\mathcal{T}\bar{M}_y)^2= \textrm{e}^ {{\rm{i}} 2 \pi q_z}$ ($q_z$ in units of the reciprocal lattice vector $2\pi/c$). On the (100) surface, both $\textit{q}_z=0$ and $\textit{q}_z=0.5$ are $\mathcal{T}\bar{M}_y$ invariant lines, and $(\mathcal{T}\bar{M}_y)^2=-1$ only occurs when $\textit{q}_z=0.5$, corresponding to the $\bar{\textrm{B}}$-$\bar{\textrm{D}}$ high-symmetry line. As a result, the doubly degenerate surface nodal line along $\bar{\textrm{B}}$-$\bar{\textrm{D}}$ is protected by the $\mathcal{T}\bar{M}_y$ symmetry.

\begin{figure*}
\centering
\includegraphics[width=0.8\linewidth]{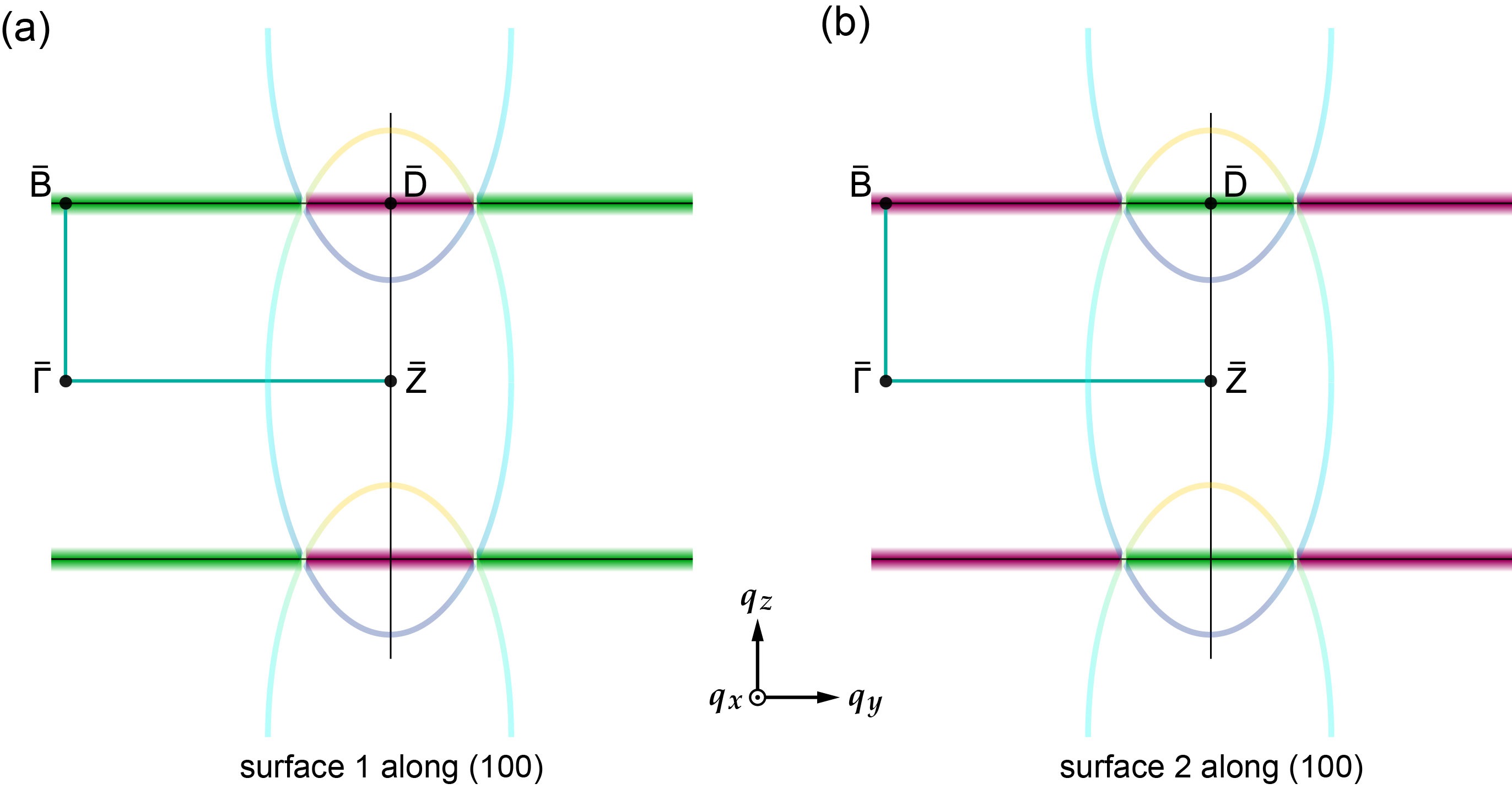}
\caption{Zak phase analysis. Zak phase of (a) surface 1 and (b) surface 2 along the (100) direction. The purple and green shaded areas correspond to the total Zak phase $\gamma = 2\pi$ and $\gamma = 0$ respectively.
}
\label{Zak} 
\end{figure*}

If we change the surface termination (denoted as surface 2) along the (100) direction, the doubly degenerate surface states along the $\bar{\textrm{B}}$-$\bar{\textrm{D}}$ high-symmetry line in Figure~\ref{surface}(f) no longer exist in the frequency range 6.410-6.415 THz, but move to the lower frequency range 6.384-6.410 THz. The distribution of the surface nodal line changes to the other side of $\bar{\textrm{B}}$-$\bar{\textrm{D}}$, from the purple shaded area in Figure~\ref{Zak}(a) to that in Figure~\ref{Zak}(b).

To explain the alternation between the distributions of the surface nodal line, we need to understand how $\mathcal{T}\bar{M}_y$ influences the Zak phase on the (100) surface Brillouin zone. The Zak phase provides information of the distribution of the doubly degenerate surface states on the (100) surface, %As the drumhead-like surface states originate from the $\pi$ Berry phase of the nodal ring in SnIP, we can also use the Zak phase $\gamma$ to explain their existence and evolution on the (100) surface. 
and can be integrated along the (100) direction, 
\begin{equation}
    \gamma(\textit{q}_b,\textit{q}_c)={\rm{i}} 
    \int_{-\pi}^{\pi}\sum_{\nu={\rm occu}.} 
    \bra{ \psi(\textbf{q}\nu) }
    \frac{\partial}{\partial \textit{q}_a}     
    \ket{ \psi(\textbf{q}\nu)     }
    d \textit{q}_a,
\end{equation}
where $\psi(\textbf{q}\nu)$ is the phonon eigenvector with wave vector \textbf{q} and band index $\nu$. Note that on the (100) surface Brillouin zone, $\textit{q}_b \parallel \textit{q}_y$ and $\textit{q}_c \parallel \textit{q}_z$, therefore we use ($\textit{q}_x$,$\textit{q}_y$,$\textit{q}_z$) instead of ($\textit{q}_a$,$\textit{q}_b$,$\textit{q}_c$) to be consistent with $\mathcal{T}\bar{M}_y$. %, as $\textit{q}_b // \textit{q}_y$ and $\textit{q}_c // \textit{q}_z$. 
When $\textit{q}_z=0.5$, $(\mathcal{T}\bar{M}_y)^{2}=-1$ and $\psi_1 (-\textit{q}_x,\textit{q}_y,0.5) = (\mathcal{T}\bar{M}_y) \cdot \psi_2 (\textit{q}_x,\textit{q}_y,0.5)$. Therefore, we can classify the phonon states at constant $\textit{q}_y$ and $\textit{q}_z = 0.5$ into two subgroups $\{\psi_1 (\textbf{q}) \}$ and $\{\psi_2 (\textbf{q}) \}$,  %when $\textit{q}_a = 0$ or 0.5, 
\textit{i.e.}, $\{\psi_1 (\textbf{q}) \}$ and $\{\psi_2 (\textbf{q}) \}$ are $\mathcal{T}\bar{M}_y$ related. Therefore, for the $\bar{\textrm{B}}$-$\bar{\textrm{D}}$ high-symmetry line ($\textit{q}_z=0.5$) on the (100) surface Brillouin zone, the Zak phases for the two subgroups $\{\psi_1 (\textbf{q}) \}$ and $\{\psi_2 (\textbf{q}) \}$ are equal to each other, \textit{i.e.} $\gamma_1 (\textit{q}_y,0.5) = \gamma_2 (\textit{q}_y,0.5)$. Due to $\mathcal{PT}$, $\gamma_1 (\textit{q}_y,0.5) = \gamma_2 (\textit{q}_y,0.5)$ is either 0 or $\pi$, which depends on the surface termination (\textit{i.e.} choice of the inversion center). When $\gamma_1 (\textit{q}_y,0.5) = \gamma_2 (\textit{q}_y,0.5) = 0$, there are no surface states as the total Zak phase $\gamma = \gamma_1 + \gamma_2 = 0$. When $\gamma_1 (\textit{q}_y,0.5) = \gamma_2 (\textit{q}_y,0.5) = \pi$, the total Zak phase $\gamma = 2\pi$, leading to doubly degenerate surface states. For surface 1, the purple shaded line in Figure~\ref{Zak}(a) has a total Zak phase of 2$\pi$ with two surface states distributed in this area, while the green shaded area corresponds to $\gamma = 0$ and there are no surface states in this area. 
%The obtained surface Brillouin zone for surface 1 can be divided into three parts according to $\gamma = 2\pi,\pi,0$, corresponding to the purple shaded area, the cyan shaded area and the remaining parts without any colour in Figure~\ref{Zak}(a). With $\gamma=2\pi$, there are two surface states in the purple shaded area. In the cyan shaded area, we have $\gamma=\pi$, which gives rise to one surface state. For the remaining parts without any colours, $\gamma=0$, and there are no surface states in this area on the (100) surface. 
% We note that the Zak phase is a gauge variant and a mod 2 topological invariant, so the states with $\gamma=0$ and $\gamma=2\pi$ can be modulated by changing the termination of the material, which can lead to either zero or two surface arcs in the purple shaded area.
% \sout{The purple shaded area is the most interesting one, as $\gamma$ can be varied from 0 to 2 once the termination of SnIP is changed, which can lead to either zero or two surface arcs in the purple shaded area.} 
%This is because the Zak phase of the purple shaded area and the colourless area swaps, as shown in Figure~\ref{Zak}(b).
%Therefore, we can get the surface nodal line on the (100) surface along $\bar{\textrm{B}}$-$\bar{\textrm{D}}$ in the purple shaded area either in Figure~\ref{Zak}(a) or in Figure~\ref{Zak}(b).
For surface 2, the Zak phase of the purple shaded area and the colourless area swap, as shown in Figure~\ref{Zak}(b), and the distribution of the surface nodal line switches to the new purple shaded area. It should be noticed that the standard definition of Zak phase is in terms of modulo 2$\pi$, and a Zak phase of 2$\pi$ is therefore equal to zero under this definition. However, in our case, a total Zak phase of 2$\pi$ corresponds to $\gamma_{1}=\gamma_{2}=\pi$, \textit{i.e.} two $\pi$ Zak phase within each subgroup, indicating two surface states emerging from each subgroup. On the other hand, a total Zak phase of zero corresponds to two zero Zak phase within each group, leading to vanishing surface states.

%, as shown in Figure~\ref{TMy}(b). 
%, indicating that, for $\textit{q}_y = \textrm{constant}$ and $\textit{q}_z = 0.5$, $\psi_1 (\textbf{q})$ and $\psi_2 (\textbf{q})$ are mutually symmetric with respect to $\textit{q}_x = 0$.
%At $\textit{q}_a = 0$ or 0.5, $\psi_1 (\textbf{q})$ and $\psi_2 (\textbf{q})$ are degenerate; apart from $\textit{q}_a = 0$ and 0.5, $\psi_1 (\textbf{q})$ and $\psi_2 (\textbf{q})$ are symmetric with respect to $\textit{q}_a = 0$, as confirmed by phonon bands 91-94 in Figure~\ref{TMy}(a).
%\textit{e.g.} at 6.400 THz in Figure~\ref{surface}(g), where the isofrequency contours exhibit doubly degenerate surface states.
%\sout{the Zak phase is different from the top one because the termination changes. As shown in the surface states in Figure~\ref{surface}(f), along the $\bar{\textrm{B}}$-$\bar{\textrm{D}}$ high-symmetry line, the doubly degenerate surface nodal line no longer exists in 6.410-6.415 THz, but moves down to 6.384-6.410 THz. Here we only show one example at 6.400 THz in Figure~\ref{surface}(g), where the isofrequency contours exhibit doubly degenerate surface states.}

%\begin{figure*}[h]
%\centering
%\includegraphics[width=0.85\linewidth]{TMy.jpg}
%\caption{
%(a) Phonon bands 91-94 along the A'-B-A high-symmetry lines. (b) Schematic of phonon bands along \textit{\textbf{q}}$_a$ at $\textit{q}_b = \textrm{constant}$ and $\textit{q}_c = 0.5$.
%}
%\label{TMy} 
%\end{figure*}

It is worth mentioning that bending SnIP rods does not break the time-reversal and glide mirror symmetries on the (100) surface of SnIP. Therefore, the degenerate topological surface states remain unchanged under bending. % Experimental measurements have demonstrated that SnIP is an ultrasoft material, and bending this material up to a degree of 90$^{\circ}$ does not change the Raman frequencies. Therefore, we can have robust topological properties for both the bulk and the surface states in the phonon dispersion of SnIP, and such topological features will not disappear even under large bending.

% Similar to the (001) surface, the LDOS for both the bulk and surface states on the (010) surface is also confined in the middle of the surface Brillouin zone. Only on the (100) surface can we obtain long surface arcs in the whole Brillouin zone. Therefore, the surface states on the plane of the bulk nodal rings, \textit{i.e.}, the (100) plane, have much larger influence than those perpendicular to the bulk nodal rings.

\section*{Discussion}

In terms of fundamental physics and applications, topological bulk phonons in double helical SnIP may lead to non-dissipative surface phonons, which offer a promising platform for designing thermal devices. The existence of surface nodal lines also provides opportunities to study surface transport properties for thermoelectric devices. As another example, the bulk and surface nodal lines/rings in the phonon spectra result in large phonon DOS, which may also enhance surface superconductivity. Additionally, because of its quasi-one-dimensional crystal structure, SnIP could also be a promising material platform to investigate one-dimensional physical models such as the Su-Schrieffer-Heeger (SSH) model \cite{Su1980} and  Majorana quantum wires \cite{Kitaev2001,Franz2013}. With its high mechanical flexibility, it is also interesting to investigate the interplay between flexoelectricity and topology. Our work provides a starting point to explore all these phenomena and applications.

%the surface states of topological phonons coupling with electrons could significantly impact superconductivity: surface phonons, as well as disorder, can enhance $T_{c}$ in thin films [\textit{Phys. Rev. Lett.} 21, 1320 (1968); \textit{Phys. Rev. Lett.} 21, 1441 (1968); \textit{Phys. Rev. B} 7, 3028 (1973)]. In a similar fashion, the large density of states that topology-driven degeneracies cause in the phonon spectrum could provide a mechanism to affect superconducting critical temperatures. 

%\section*{Conclusion}

In summary, we find that double helical SnIP exhibits nodal-ring phonons in the bulk and nodal-line phonons on the surface based on first-principles calculations and group theory analysis. Benefiting from a quasi-one-dimensional crystal structure, the lattice vibrational modes of the intrachain phosphorus atoms along the helices are highly dispersive. Due to the winding of the helix P chain, these phonon bands are folded at the Brillouin zone boundary, forming single nodal rings across the whole Brillouin zone. The overlap between the two neighbouring nodal rings on the (100) surface leads to a surface nodal line that is protected by the combination of time-reversal and glide mirror symmetries $\mathcal{T}\bar{M}_y$. We also propose that different surface terminations result in different distributions of the surface nodal line, which can be understood by different Zak phases in the presence of $\mathcal{T}\bar{M}_y$. We demonstrate that, similar to the bulk nodal lines, the surface nodal lines can widely exist when protected by extra symmetries. Our discoveries may offer opportunities to study the degeneracy of the topological surface states and their transport properties.

\section*{Methods}

\subsection*{Density functional theory calculations}

All density functional theory (DFT) calculations are performed using the Vienna \textit{ab-initio} simulation package ({\sc vasp}) \cite{Kresse1996,Kresse1996a}. The generalized gradient approximation (GGA) is used in the Perdew-Burke-Ernzerhof (PBE) parametrization for the exchange-correlation functional with 14 valence electrons ($4d^{10}5s^25p^2$) for Sn, 7 valence electrons ($5s^25p^5$) for I, and 5 valence electrons ($3s^23p^3$) for P. We use a plane-wave basis with a kinetic energy cutoff of 600 eV and a 7$\times$5$\times$3 \textbf{k}-mesh for structural relaxation until the energy difference is lower than 10$^{-6}$ eV and the Hellman-Feynman force difference is lower than 10$^{-4}$ eV \AA$^{-1}$. The D2 method of Grimme is used to describe the van der Waals interactions between the double helices \cite{Grimme2006}.

\subsection*{Lattice dynamics properties}

The Hessian matrix and phonon frequencies are calculated based on the finite difference method using a 2$\times$2$\times$1 supercell with a 3$\times$3$\times$3 \textbf{k}-mesh. The phonon dispersion is obtained with the {\sc phonopy} code \cite{Togo2008,Togo2015}. The convergence of the supercell is checked by comparing the 2$\times$2$\times$1 and 1$\times$1$\times$2 supercells in Supplementary Note 1 of the Supplementary Information. The Born effective charges are also computed using a perturbative approach to account for the splitting between the longitudinal and transverse optical phonon modes (LO-TO splitting) near the Brillouin zone center \cite{Gajdos2006}, and we find that the LO-TO splitting has a minor influence on the topological properties of phonons because the topological features are far away from the zone center. 

\subsection*{Topological properties}

To obtain the topological properties of the lattice vibrational modes, we use {\sc WannierTools} to calculate the distribution of the band crossing points/lines in the whole Brillouin zone \cite{Wu2018}. The phonon surface states on the (100) surface are computed using the surface Green's function.


\section*{Data Availability} 

The datasets generated during and/or analysed during the current study are available from the corresponding authors on reasonable request.

\section*{Code Availability} 

The related codes are available from the corresponding authors on reasonable request.

\section*{Acknowledgements}

B.P. and B.M. acknowledge support from the Winton Programme for the Physics of Sustainability, and B.M. also acknowledges support from the Gianna Angelopoulos Programme for Science, Technology, and Innovation.
S.M. and T.Z. acknowledge the supports from JSPS KAKENHI Grants No.\,JP18H03678 and No.\,JP20H04633, Tokodai Institute for Element Strategy (TIES) funded by MEXT Elements Strategy Initiative to Form Core Research Center. T.Z. also acknowledge the support by Japan Society for the Promotion of Science (JSPS), KAKENHI Grant No.\,JP21K13865.
The calculations are performed using resources provided by the Cambridge Tier-2 system operated by the University of Cambridge Research Computing Service (http://www.hpc.cam.ac.uk) and funded by EPSRC Tier-2 capital grant EP/P020259/1, and also with computational support from the UK Materials and Molecular Modelling Hub, which is partially funded by EPSRC (EP/P020194), for which access is obtained via the UKCP consortium and funded by EPSRC grant EP/P022561/1.

\section*{Author Contributions} 

B.P., B.M. and T.Z. devised the project idea. B.P. performed the first-principles calculations. S.M. and T.Z. analysed the topological features. B.P., B.M. and T.Z. prepared the main part of the manuscript. B.P., S.M., B.M. and T.Z. discussed the results and the ideas for analysis and edited the manuscript.

\section*{Competing Interests}

The Authors declare no Competing Financial or Non-Financial Interests.

\end{document}